\documentclass[aps,prl,twocolumn,showpacs,preprintnumbers,amsmath,amssymb,superscriptaddress]{revtex4-1}
\usepackage{fullpage,amsthm,amsmath,amsfonts,bm,times,color,bbm}
\usepackage{graphicx,subfigure}
\usepackage[english]{babel}
\usepackage{mathtools}
\usepackage{multirow}
\usepackage{cases}
\usepackage{float}

\def\eqa{\begin{eqnarray}}
\def\eea{\end{eqnarray}}
\newcommand{\eq}{\begin{equation}}
\newcommand{\ee}{\end{equation}}

\makeatother

\makeatletter

\begin{document}

\title{Possible Origin of Memory in Earthquakes:
 Real catalogs and ETAS model % model and real networks
 }

\author{Jingfang Fan}
%\email{j.fang.fan@gmail.com}
\affiliation{Blaustein Institutes for Desert Research, Ben-Gurion University of the Negev, Midreshet Ben-Gurion 84990, Israel}
\affiliation{Department of Physics, Bar Ilan University, Ramat Gan 52900, Israel}

\author{Dong Zhou}
%\email{zhoudongbnu@gmail.com}
\affiliation{Blaustein Institutes for Desert Research, Ben-Gurion University of the Negev, Midreshet Ben-Gurion 84990, Israel}
\affiliation{Department of Physics, Bar Ilan University, Ramat Gan 52900, Israel}
\affiliation{School of Reliability and Systems Engineering, Beihang University, Beijing, 100191, China}

\author{Louis M. Shekhtman}
%\email{lsheks@gmail.com}
\affiliation{Department of Physics, Bar Ilan University, Ramat Gan 52900, Israel}

\author{Avi Shapira}
%\email{***}
\affiliation{Institute for Regulation of Emergency and Disaster, Israel}

\author{Rami Hofstetter}
%\email{***}
\affiliation{Geophysical Institute of Israel, Israel}

\author{Warner Marzocchi}
%\email{***}
\affiliation{Istituto Nazionale di Geofisica e Vulcanologia, Rome, Italy}

\author{Yosef Ashkenazy}
%\email{ashkena@bgu.ac.il}
\affiliation{Blaustein Institutes for Desert Research, Ben-Gurion University of the Negev, Midreshet Ben-Gurion 84990, Israel}

\author{Shlomo Havlin}
%\email{havlin@ophir.ph.biu.ac.il}
\affiliation{Department of Physics, Bar Ilan University, Ramat Gan 52900, Israel}

\begin{abstract}
  Earthquakes are one of the most devastating natural disasters that plague society. A skilled, reliable earthquake forecasting remains the ultimate goal for seismologists. Using the detrended fluctuation analysis (DFA) and conditional probability (CP) methods we find that memory exists not only in inter-occurrence seismic records, but also in released energy as well as in the series of the number of events per unit time. Analysis of the conventional earthquake model (Epidemic Type Aftershock Sequences, ETAS) indicates that earthquake memory can be reproduced only for a narrow range of model's parameters. This finding, therefore provides additional accuracy on the model parameters through tight restrictions on their values in different worldwide regions and can serve as a testbed for existing earthquake forecasting models.
  %Here we utilize detrended fluctuation analysis (DFA) and conditional probability (CP) methods to study memory in both real seismic catalogs and the Epidemic Type Aftershock Sequences (ETAS) model. We find that memory exists not only in inter-occurrence seismic records, but also in the times series of the number of events and released energy. Furthermore, we analyze how the ETAS model's parameters affect the memory and find that ETAS is able to reproduce memory similar to that of real catalogs, but only with specific parameters values that differ from previous estimates. We also consider the origin of the memory, and find that the background rate $\mu$ in the model affects the memory through interference of temporally overlapping aftershock subsequences. The exponent relating the production of aftershocks as a function of magnitude, $\alpha_M$, and the power $p$ from the Omoris law can also affect memory through the branching ratio of the ETAS model.
%The framework presented here not only shows the importance of memory in real seismic catalogs and models, but also facilitates the improvements of earthquake forecasting. 
\end{abstract}
\date{\today}

%\flushbottom
\maketitle

The process through which earthquakes occur is complex involving spatio-temporal dynamics \cite{sornette_earthquakes:_1999,sornette_notitle_2006} and has previously been characterized as a paradigm of self-organized criticality \cite{bak_unified_2002,sammis_positive_2002}. However, the underlying mechanisms of earthquakes are still not fully understood \cite{turcotte1997fractals}, and as a consequence, predicting events' magnitude, location, and time in advance remains elusive. Over the past decades, scaling laws for the distribution of waiting times between earthquake events have been obtained in seismic data \cite{bak_unified_2002,corral_long-term_2004} as well as in rock fracture experiments in laboratories \cite{davidsen_scaling_2007}. Perhaps the most promising observation is that a rescaling involving region size and magnitude threshold, produces data collapse onto a universal gamma distribution  for many worldwide regions \cite{corral_long-term_2004}. This observation is of great importance for the development of physical and statistical models of earthquake dynamics. However, recent analysis on the Epidemic-Type Aftershock Sequences (ETAS) model, have indicated that the distribution is not universal \cite{molchan_spatial_2005,saichev_universal_2006}, but instead it is fundamentally a bimodal mixture distribution \cite{touati_origin_2009}. These modeling studies have captured much of the earthquake dynamics through the distribution of recurrence intervals but they have not considered the memory found in real earthquakes time series. 
%Here we consider the ability of the ETAS model to reproduce thememory observed in earthquake data.
%and assess the impact that adding memory has on the model.

Long-term memory has been reported in many natural systems, such as climate systems \cite{noauthor_robustness_1969,koscielny-bunde_analysis_1996,koscielny-bunde_indication_1998}, physiology \cite{peng_long-range_1993,bunde_correlated_2000}, and even, as mentioned above, in seismic activity \cite{livina_memory_2005,lennartz_long-term_2008}. Indeed Livina \textit{et. al} found that consecutive recurrence times (for different magnitude threshold) depend on each other, such that each short and long recurrence times tend to cluster in time (i.e., short after short and long after long) \cite{livina_memory_2005}. Later, Lennartz \textit{et. al} studied the Northern and Southern California earthquake catalogs and found long-term memory using  DFA. The goal of the present study is to uncover the mechanisms that underlie the memory observed in earthquake data. This is done by analyzing the ETAS model and by assessing what parameter values are able to capture the observed memory. In this Letter, we utilize the detrended fluctuation analysis (DFA) and conditional probability (CP) methods to analyze the memory in two highly  accurate real seismic catalogs (Italian and Israeli) and in the ETAS model. We find that the origin of memory in the ETAS model is influenced by (i) the background rate $\mu$ in the model which affects the memory through interference of temporally overlapping aftershock subsequences, i.e., smaller $\mu$ leads to stronger memory; (ii) the exponent relating the production of aftershocks as a function of magnitude, $\alpha_M$, and the power $p$ of the Omori's law can also affect memory through the branching ratio of the ETAS model, i.e., smaller $p$ and larger $\alpha_M$ result in stronger memory. Analysis of the ETAS model indicates that earthquake memory can be reproduced only for a narrow range of model's parameters.% Still little is known about the origin of the memory in earthquakes.

% Our approach can also be used to improve the forecasting ability of other related models. [THIS SENTENCE DOES NOT BELONG HERE.]
DFA is a well-established method for the detection of long-range correlations in time series \cite{peng_mosaic_1994-1}. It has successfully been applied to many fields, such as DNA \cite{buldyrev_long-range_1995}, heart rate dynamics \cite{peng_long-range_1993,ashkenazy_discrimination_1998,bunde_correlated_2000}, climate records \cite{koscielny-bunde_indication_1998}, and others. If the data possess long term correlations, the fluctuation function,
% {\color{red} LMS: Can we perhaps define $F(n)$ too? How is it found?}
$F(n)$, increases according to a power-law relation: $F(n) \sim n^\alpha$ where $n$ is the window size and $\alpha$ is the scaling exponent. The exponent $\alpha$ is calculated as the slope of a linear fit to the log-log graph of $F$ vs $n$. An exponent $\alpha=0.5$ indicates that there are no long-range correlations (white noise), whereas $\alpha >0.5$ indicates that the record posses long term positive correlations (higher values of $\alpha$ imply stronger correlations). 

\begin{figure}
\begin{centering}
\includegraphics[width=1.0\linewidth]{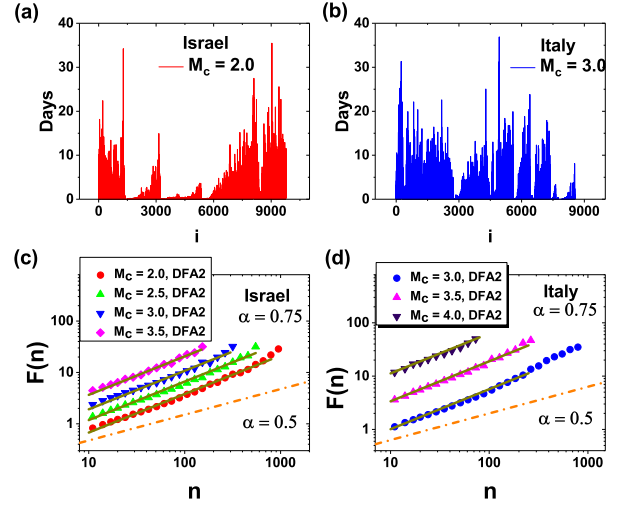}
\caption{\label{Fig1} Interoccurrence time series between earthquake events from the Israeli (a) and Italian (b) catalogs with magnitude threshold $M_c$. Detrended fluctuation analysis \cite{peng_mosaic_1994-1} of the interoccurrence times from the (c) Israeli and (d) Italian catalogs. The solid lines are the best fit lines with slope $\alpha \approx 0.75$ with R-square $>0.99$. For comparison we also show a dashed dotted line that indicates no memory (dashed line, $\alpha=0.5$). }
\end{centering}
\end{figure}

\begin{figure}
\begin{centering}
\includegraphics[width=1\linewidth]{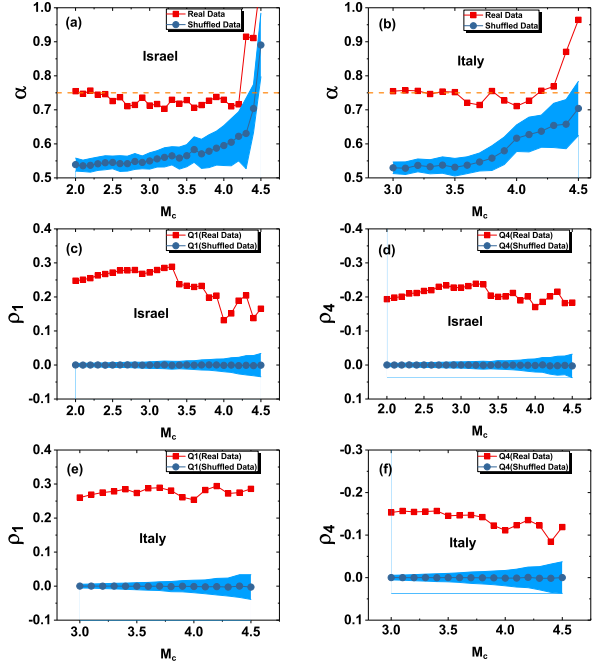}
\caption{\label{Fig2} DFA memory scaling exponent $\alpha$ as a function of magnitude threshold $M_c$ for the (a) Israeli and (b) Italian catalogs. CP memory coefficient $\rho1$ and $\rho4$ as a function of magnitude threshold $M_c$ for (c-d) Israeli and (e-f) Italian catalogs. The blue lines in the bottom of each subplot indicate the DFA scaling exponent $\alpha$ and the CP memory measure $\rho1$, $\rho4$ for the control randomly shuffled records; the shaded region indicates the error bars.
}
\end{centering}
\end{figure}

Here, we study the long-term memory in the real seismic catalogs of Italy and Israel \cite{schorlemmer_completeness_2010,gasperini_empirical_2013,DATA_ISRAEL_ITALY}.  Fig.~\ref{Fig1} shows the results on examples of the return intervals in the  Israeli (from 1981 to 2017) and Italian (from 1986 to 2017) seismic catalogs with different magnitude thresholds, $M_c$. One should note that, the earthquake events with $M \geq 2.0$ $(3.0)$ for Israeli (Italian) are complete, meaning that all earthquakes above this magnitude are included in the catalog. The DFA results on the inter-occurrence time series  are presented in Fig.~\ref{Fig1} (c, d) and indicate the existence of similar long-term memory in both the Israeli and Italian earthquake catalogs. These results suggest that the scaling exponent is close to $\alpha = 0.75$ and is independent of the magnitude $M_c$ [see, Fig.~\ref{Fig2} (a) and (b)]. The  values of $\alpha$ found in our study for Israel and Italy are consistent with the previous study of the southern and northern California catalogs \cite{lennartz_long-term_2008}. To demonstrate that this memory is not accidental, we analyzed also the randomly reshuffled earthquake catalog records. The shuffling procedure destroys the correlations between the return intervals but keeps the distribution of the return intervals. We considered 1000 such shuffled records and determined the averaged $\alpha \pm$ standard deviation, as shown in Fig.~\ref{Fig2} (a) and (b). Compared with the shuffled data, we find that real data exhibits significant memory and the exponents of the data are significantly larger than the exponents of the shuffled data. The large increase of the long-term memory in both real and shuffled data  for large $M_c$ is probably due to finite size effects \cite{eichner_extreme_2006}. 

We also performed DFA on other seismic variables, such as the number of earthquake events and released energy within a coarseness time window $dt$; to our knowledge such analysis have not been previously reported.
% Such variables provide more information, e.g., number of mainshocks or aftershocks and the corresponding magnitudes.
For each catalog we define a time series, $S(t)$, where each term is related to the earthquake activity that takes place within the time window, $dt$, \cite{tenenbaum_earthquake_2012}, $S(t) = \sum_{l=1}^{E(t)} 10^{\frac{3}{2}M_{l}(t)},$
% \begin{equation}
% S(t) = \sum_{l=1}^{E(t)} 10^{\frac{3}{2}M_{l}(t)},
% \label{eq1}
% \end{equation}
where $E(t)$ denotes the number of events that occurred between $t$ and $t+dt$. The signal is proportional to the total energy released in a $dt$ time period. The filtered record is then defined as, $s(t)=\log(S(t))$ for $S(t) > 1$ and zero for $S(t) \leq 1$ and $e(t)=\log(E(t))$ for $E(t) > 1$ and zero for $E(t) \leq 1$.
% \begin{numcases}{s(t)=}
%    \log(S(t)),  & for $S(t) > 1$; \nonumber\\ 
%    0, & for  $S(t) \leq 1$. \nonumber
% \end{numcases}
% \begin{numcases}{e(t)=}
%    \log(E(t)),  & for $E(t) > 1$; \nonumber\\
%    0, & for  $E(t) \leq 1$.
% \end{numcases}
This (log) operation aims to suppress extremely high values of $E(t)$ and $S(t)$ that can affect $F(n)$ and its correlation exponent $\alpha$. Fig.~\ref{Fig:S2} and \ref{Fig:S3} depict the DFA analysis for $s(t)$ and $e(t)$ for the Israeli and Italian catalogs, for different time windows $dt$. We find that for both countries and for all studied magnitudes  the value of $\alpha$ is quite robust $\sim 0.75$. This indicates that return intervals, the number of events and released energy are significantly correlated and follow the same scaling exponent. This, apparent universal scaling exponent, $\alpha$, can potentially be used to validate the performance of earthquake  forecasting models and for narrowing the range of model parameters.

According to Omori's law \cite{omori1894after} earthquake events tend to cluster in time due to the time-dependent relaxation of the crust through the release of triggered aftershocks. The rate, $n(t)$, of aftershocks above a certain magnitude $M_c$ decays with time $t$ as $n(t) \sim t ^{-p}$. This clustering and power-law decay indicates that both short and long- term correlations (memories) exist in seismic data. However, the aforementioned DFA can only quantify the long term memory ($\alpha>0.5$), and thus to better characterize and understand all types of memory in earthquake events, we now further develop  and apply a general CP method. 
%[NOTE THAT BOTH THE DFA AND CP WERE PUBLISHED IN THE PAST AND WE SHOULD NOT STRESS THEM TOO MUCH AS THE REVIEWERS MAY COMPLAIN THAT OUR WORK IS NOT ORIGINAL... I SUGGEST TO REMOVE SOME OF THE FIGURES OF THE DFA AND CP ANALYSIS AND TO PROVIDE MORE RESULTS REGARDING THE MODEL.]

\begin{figure}
\begin{centering}
\includegraphics[width=1\linewidth]{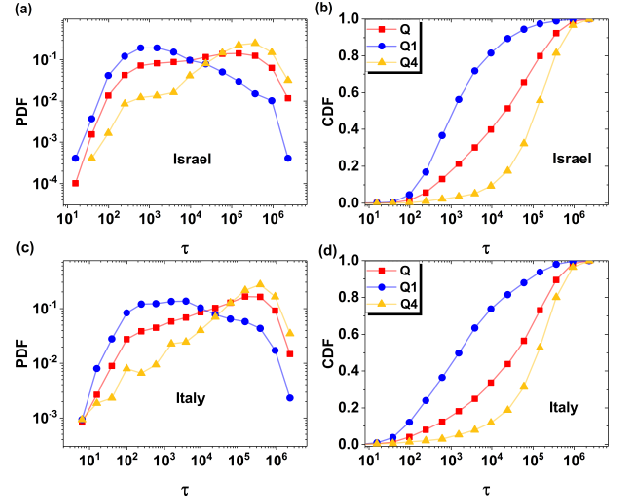}
\caption{\label{Fig3} (a,c) Conditional PDF and (b,d) CDF of the recurrence times $\tau$ for the (a,b) Israeli catalog above the threshold $M_c = 2.0$ and for the (c,d) Italian catalog above the threshold $M_c = 3.0$.}
\end{centering}
\end{figure}

We begin by sorting the full time series of recurrence intervals in ascending order and divide it into four 25\% quantiles; i.e., the first quantile, Q1, represents shortest $25\%$ of waiting times, etc. We will next consider the distribution of recurrence times, $\tau$, that follow a prior recurrence time $\tau_0$, $P(\tau|\tau_0)$, where $\tau_0$ belongs to either one of the quantiles at the extremities, Q1 or Q4. Essentially, given that a prior recurrence time was either short (in Q1) or long (in Q4), we ask what is the distribution of the subsequent recurrence times. In records without memory, $P(\tau|\tau_0)$ should not depend on $\tau_0$ and should be identical to $P(\tau)$. 
%Since multiple time scales are involved in the analysis (from seconds to many years), we define bins over the probability density of the recurrence time which are exponentially growing, $2.5^q $, with $q$ labeling consecutive bins \cite{corral_long-term_2004}. [THE LAST SENTENCE IS NOT TOTALLY CLEAR.]
Figs.~\ref{Fig3} (a) and (c) show the PDF of waiting times [Q, Q1 and Q4] for the Israeli and Italian catalogs, respectively. The figure suggests, as in Refs.~\cite{livina_memory_2005}, that $P(\tau|\tau_0)$ depends strongly on the previous recurrence time $\tau_0$, such that short recurrence times are more likely to be followed by short ones, and long recurrence times follow long ones. Note that, the present study is different from the analysis of  Refs.~\cite{corral_long-term_2004,livina_memory_2005}, as the distribution of recurrence times in our analysis is not rescaled and normalized by the mean event rate.

We next consider the Cumulative Distribution Function (CDF) of the recurrence times and quantify the difference between the overall CDF of the unconditional recurrence times for the entire catalog, Q, and the CDF for a given quantile by considering the area represented in the gap between the curves (see Fig.~\ref{Fig3} (b) and (d)). Here, we denote the CDF of the recurrence time for Q, Q1 and Q4 as $CQ(\tau)$, $CQ1(\tau)$ and $CQ4(\tau)$, respectively. To this end, we define the level of memory for Q1 as, $\rho1 = \int(CQ1(\tau) - CQ(\tau)) d\tau/{\int d\tau}$,  
% \begin{equation}
% \rho = \frac{\int(Q1(\tau) - Q(\tau)) d\tau }{\int d\tau},
% \end{equation}
similarly, the level of memory for Q4 as, $\rho4 = \int(CQ4(\tau) - CQ(\tau)) d\tau/{\int d\tau}$. Thus, $0\leq\rho1\leq 1$ and $-1\leq\rho4\leq 0$,
%
% the Jaccard index (Intersection over Union), $\rho = \frac{S1}{S1+S2}$, ($0\leq\rho\leq$1), for Q1 and Q4, 
%  
and higher $|\rho1|$ (or $|\rho4|$), implies stronger memory and $\rho1 = 0$ (or $\rho4 = 0$), implies no memory. Fig. \ref{Fig3} shows the results for Israeli and Italian earthquake catalogs. We find $\rho1=0.248, \rho4=-0.193$ for Q1 and Q4 of the Israeli catalog whereas $\rho1=0.260, \rho4=-0.153$ for Q1 and Q4 of the Italian catalog. Fig.~\ref{Fig2} (c--f) suggest that the values of $\rho$ are robust and do not depend on $M_c$. These results are consistent with the DFA results presented above.

We will now study the possible origin of the memory in real data by analyzing the ETAS model \cite{ogata_statistical_1988} that generates synthetic catalogs. ETAS is a stochastic point-process model in which background events occur through a Poisson process in time with rate $\mu$, and all past events above a threshold magnitude $M_c$ may produce aftershocks. It has been successfully used for operational earthquake forecasting, e.g., the complex Amatrice-Norcia seismic sequence \cite{marzocchi_earthquake_2017}. The ETAS model is based on two well established empirical basic laws: (i) the Gutenberg-Richter law \cite{gutenberg_frequency_1944}, $\log N = a - bm$, where $N$ is the number of events in a given time period with magnitude $\geq m$, and $a$, $b$ are constants; (ii) the Omori law, $n(t) = K/(c+t)^p$, where $K$, $c$, and $p$ are constants and $t$ denotes the time. The conditional intensity function $\lambda$ in the ETAS model is,
\begin{equation}
\lambda (t|H_{t}) = \mu + A\sum\limits_{i:t<t_{i}} \exp[\alpha_{M}(m_{i} - M_{c})]\left(1 + \frac{t - t_{i}}{c} \right)^{-p},
\label{eq3}
\end{equation}
where $t_{i}$ are the times of the past events and $m_{i}$ are their
magnitudes; $H_{t} = \{(t_{i}, m_{i}); t_{i} <t\}$ is the history of occurrence. Here, $A = K/cp$ is the occurrence rate of earthquakes in the Omori law at zero lag \cite{touati_origin_2009}, and $\alpha_{M}$, is called the productivity parameter. 

\begin{figure*}
\begin{centering}
\includegraphics[width=1\linewidth]{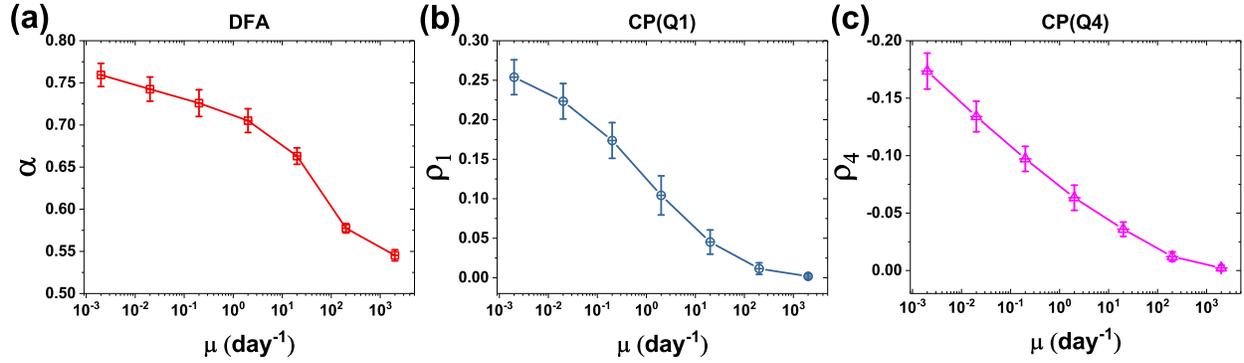}
\caption{\label{Fig4} Memory measures $\alpha$, $\rho1$ and $\rho4$ as a function of ETAS parameter $\mu$. (a) DFA scaling exponent $\alpha$, (b) CP coefficient $\rho1$ for short recurrence times $Q1$, (c) CP coefficient $\rho4$ for long recurrence times $Q4$. The other ETAS parameters used are $A = 6.26$, $\alpha_M = 1.4$, $p = 1.13$, and $c = 0.007$ for the Italian catalog, following  \cite{lombardi_estimation_2015}. The error bars are based on 100 independent simulation realizations.}
\end{centering}
\end{figure*}

We next investigate how the background rate, $\mu$, affects the memory in the ETAS model. We have only considered variations of $\mu$ without changing the branching ratio, by setting $A = 6.26$, $c = 0.007$, $\alpha_M = 1.4$ and $p = 1.13$. [These are the prior estimates for the Italian catalog, and $\mu = 0.2$ \cite{lombardi_estimation_2015}.] Fig.~\ref{Fig4} shows that the memory coefficients $\alpha$ (of the DFA) and $\rho$ (of the CP for Q1 and Q4) decay with $\mu$, which suggests that $\mu$ significantly affects the memory such that smaller $\mu$, implies stronger memory. This is because variation in $\mu$ arises from the effect of the interference of temporally overlapping aftershock subsequences or ``correlated inter-event times'' \cite{touati_origin_2009}. As $\mu$ is increased, fewer aftershocks occur and more overlapping aftershock sequences take place, increasing the fraction of independent inter-event times. Thus, the memory is destroyed. We also notice that decreasing $\mu$ affects the distribution of recurrence times, changing it from a unimodal to a bimodal distribution \cite{touati_origin_2009}.

\begin{figure*}
\begin{centering}
\includegraphics[width=1\linewidth]{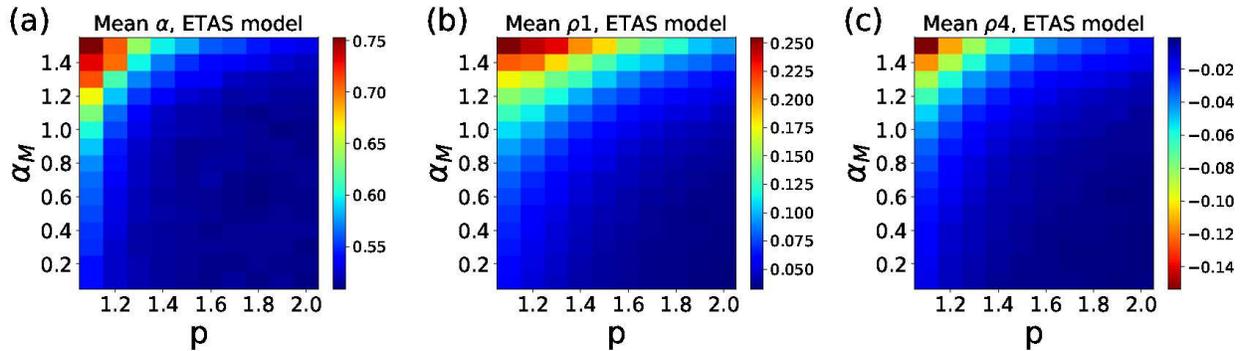}
\caption{\label{Fig5} The mean values of the (a) DFA scaling exponent $\alpha$, (b) CP coefficient $\rho1$ for short recurrence times $Q1$, and (c) CP coefficient $\rho4$ for long recurrence times $Q4$ as a function of ETAS model parameters, $p$ and $\alpha_M$. The other ETAS parameters used were $A = 6.26$, $\mu = 0.2$ and $c = 0.007$, for the Italian catalog, following \cite{lombardi_estimation_2015}.}
\end{centering}
\end{figure*}

To better understand the origin of memory in the ETAS model, we also analyzed the effects on memory of the productivity parameter $\alpha_M$ and the Omori's law power $p$. Here we chose $\mu = 0.2$ [following the prior estimates for the Italian catalog \cite{lombardi_estimation_2015}] where the other ETAS model parameters are the same as mentioned above. Fig.~\ref{Fig5} presents simulation results of the model, for $p \in [1.1, 2.0]$ and $\alpha_M \in [0.1, 1.5]$. We find that the memory in ETAS model strongly depends on both $p$ and $\alpha_M$: smaller $p$ and larger $\alpha_M$ imply stronger memory. Note that memory patterns for DFA ($\alpha$) and CP ($\rho1$, $\rho4$) are highly consistent. This result can be explained by the branching ratio $n'$, the average number of aftershocks generated by each parent event. As the branching ratio grows, correlated aftershocks increase and the background fraction decreases, resulting in stronger memory. The branching ratio $n'$ is obtained by integrating $A \exp(\alpha_M m) (1+\frac{t}{c})^{-p}$ over both time and magnitude from $0$ to $\infty$, and is given by \cite{sornette_apparent_2005}: $n' = \frac{Ac}{p - 1} \frac{\beta}{\beta - \alpha_M},$
% \begin{equation}
%
%   \label{eq4}
% \end{equation}
where $\beta = b \ln(10)$, is obtained from the Gutenberg--Richter law. The condition (with $p>1$ and $\beta>\alpha_M$) for physical stability is that $n'$ should be finite and less than 1 \cite{sornette_occurrence_2002}. The parameter $\mu$ has no effect on aftershock generation. We find that $n'$ is proportional to $\alpha_M$ and $1/p$, which agrees well with our results [see Fig.~\ref{Fig5}]. To determine the sample uncertainty, we performed 100 independent simulation realizations for each combination of parameter values. The error bars (standard deviations) are shown in Fig.~\ref{Fig:S7}. We also present the DFA of the occurrence times for a specific simulation realization with ETAS parameters, $\mu=0.2$, $A=6.26$, $\alpha_M = 1.5$, $p=1.1$ and $c=0.007$, see Fig.~\ref{Fig:S4}. The model results yield  the same alpha as in real data.
 The CP analysis for the same realization is shown in Fig.~\ref{Fig:S5}. Fig.~\ref{Fig:S6} demonstrates that the values of $\alpha$, $\rho1$ and $\rho4$ are robust, do not depend on $M_c$ in the ETAS model and are similar to the finding in real catalogs.

However, Fig. \ref{Fig4} and \ref{Fig5} show that the ETAS model can reproduce the same memory as in real catalogs, only for a small range of parameter values. %For example, the previous results for the parameters fit in ETAS for the Italian catalog are: $\mu=0.2$, $A=6.26$, $\alpha_M = 1.4$, $p=1.13$ and $c=0.007$ \cite{lombardi_estimation_2015}.
% [CAN YOU INDICATE THIS CHOICE OF PARAMETERS BY X IN THE COLOR PLOT??] 
The parameters values that reproduce the memory observed in the real catalog are $\mu=0.2$, $\alpha_M = 1.5$ and $p=1.1$. These are close to, though still statistically different from the previous parameters found, which were $\mu=0.2$, $A=6.26$, $\alpha_M = 1.4$, $p=1.13$ and $c=0.007$ \cite{lombardi_estimation_2015}. %While these values are not far from the prior results, they are still outside of the previous predicted range and may have noticeable effects on the prediction of seismic events.
In estimating the ETAS model for an earthquake catalog, the ETAS parameters are commonly inverted from the data based on the point-process maximum likelihood (ML) method, by the Davidon-Fletcher-Powell algorithm \cite{ogata_statistical_1988} or by Simulated Annealing \cite{lombardi_estimation_2015}.
% The log-likelihood for a point process is given by:
% \begin{equation}
% \log L = \sum\limits_{i} \log \lambda(t_{i}|H_t) - \int \lambda(t|H_t) dt, 
% \label{eq4}
% \end{equation}
% where $\lambda$ is the conditional intensity function, see Eq.~\ref{eq3}, $t_i$ refer to times of events, and $H_t$ is the history of the process at the time $t$. Another named expectation maximization (EM) method was also developed for an accurate estimation of the ETAS model \cite{veen_estimation_2008},  which is based on the probabilistic incorporation of the branching structure. However, both ML and EM methods do not consider the memory effects.
Our results provide a narrow range of values that are capable of reproducing the memory and provide an intuition for why only these values are reasonable.
Thus, our results and methods can be used to improve the choice of the parameters of ETAS model which can potentially help to increase the forecast rate.

In summary, we have studied several seismic catalogs (recurrence times, number of events and released energy) and found long-range memory for different magnitude threshold $M_c$. We use the DFA and CP analysis methods to quantify the level of memory and long-term correlations. We study the origin of the memory in real data by using synthetic catalogs generated by the ETAS model and find that the background rate $\mu$ affects the memory through interference of temporally overlapping aftershock subsequences; while the productivity parameter $\alpha_M$ and Omori's law power $p$ affect the memory through the branching ratio $n'$. The information on the memory can be further incorporated into the algorithm to estimate the maximum likelihood parameters of ETAS model, and thus to improve the forecast rate.
%We believe that our approach developed here is unique, and it will not only apply to more earthquake clustering models, such as ETAS with spatial components \cite{ogata_space-time_1998} and the short-term earthquake probability models \cite{gerstenberger_real-time_2005}, but also be extended to study and improve our knowledge in other fields such as climate, systems biology, and financial systems.

\section*{Acknowledgements}
We thank the  Italy-Israel project OPERA, which is funded jointly by 
the Italian Ministry of foreign affairs and international cooperation, and the Israeli Ministry of science, technology, and space; the Israel-Italian collaborative project NECST, the Israel Science Foundation, ONR, Japan Science Foundation, BSF-NSF, and DTRA (Grant no. HDTRA-1-10-1-0014) for financial support.

\bibliography{MyLibrary}

%%%%%%%%%% Merge with supplemental materials %%%%%%%%%%
\pagebreak

\newpage
\appendix
\pagebreak
\widetext
\begin{center}
\textbf{\large Supplemental Materials: Origin of Memory in Earthquakes:
 Real catalogs and ETAS model}

%Jingfang Fan, Dong Zhou, Yosef Ashkenazy and Shlomo Havlin
\end{center}

\setcounter{equation}{0}
\setcounter{figure}{0}
\setcounter{table}{0}
%\makeatletter
\renewcommand{\theequation}{S\arabic{equation}}
\renewcommand{\thefigure}{S\arabic{figure}}

%\begin{equation}
%\left\{
%\begin{array}{ll}
%\boldsymbol{x}_{<}  =  \boldsymbol{z}_{<},  \\ 
%\boldsymbol{x}_{>}   =  \boldsymbol{z}_{>} \odot e^{s(\boldsymbol{z}_{<})} + t(\boldsymbol{z}_{<}), \label{eq:rnvp}
%\end{array}
%\right.
%\end{equation}
%The transformation \Eq{eq:rnvp} is easy to invert by reversing the basic arithmetical

\section{Further results}
We present here some further results not shown in the main text.

\clearpage

%\begin{figure}
%\begin{centering}
%\includegraphics[width=1.0\linewidth]{Fig3}
%\caption{\label{Fig:S1A} (a) $S(p_c,r)$ as a function of $r$ for different $M_{inter}$. (b) $\frac{\partial S(r,p)}{\partial r}$ as a function of $p_{c}-p$ for $r = 10^{-4}$ and different $M_{inter}$. The solid green lines in (a) and (b) show the slope $1/\delta = 0.055$ and $\gamma = 2.389$ respectively. }
%\end{centering}
%\end{figure}
%
%
%
%

%\begin{figure}
%\begin{centering}
%\includegraphics[width=1.0\linewidth]{FigS3}
%\caption{\label{Fig:S1} DFA memory scaling exponent $\alpha$ as a function of magnitude threshold $M_c$ for (a) Israeli and (b) Italian catalogs.}
%\end{centering}
%\end{figure}

\begin{figure}
\begin{centering}
\includegraphics[width=1.0\linewidth]{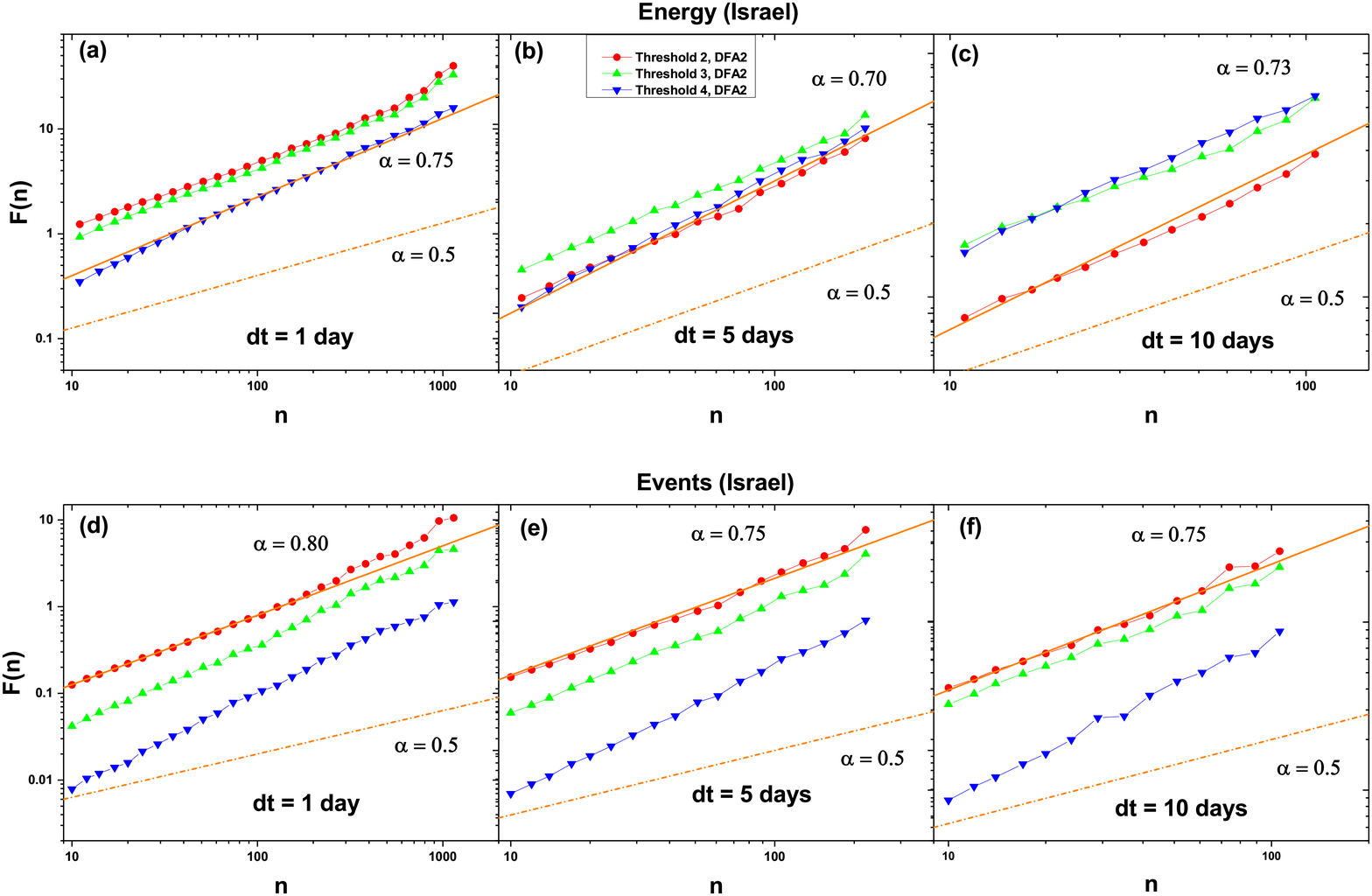}
\caption{\label{Fig:S2} DFA of (a, b and c) the released energy time series, (d, e and f) the number of events time series, defined in Eq.~2, within a $dt$ time period for  Israeli catalog. The solid lines are the best fitting lines with slope $\alpha$, R-square $>0.99$. For comparison, the dashed line with slope $\alpha=0.5$, indicating no memory, are presented.}
\end{centering}
\end{figure}

\begin{figure}
\begin{centering}
\includegraphics[width=1.0\linewidth]{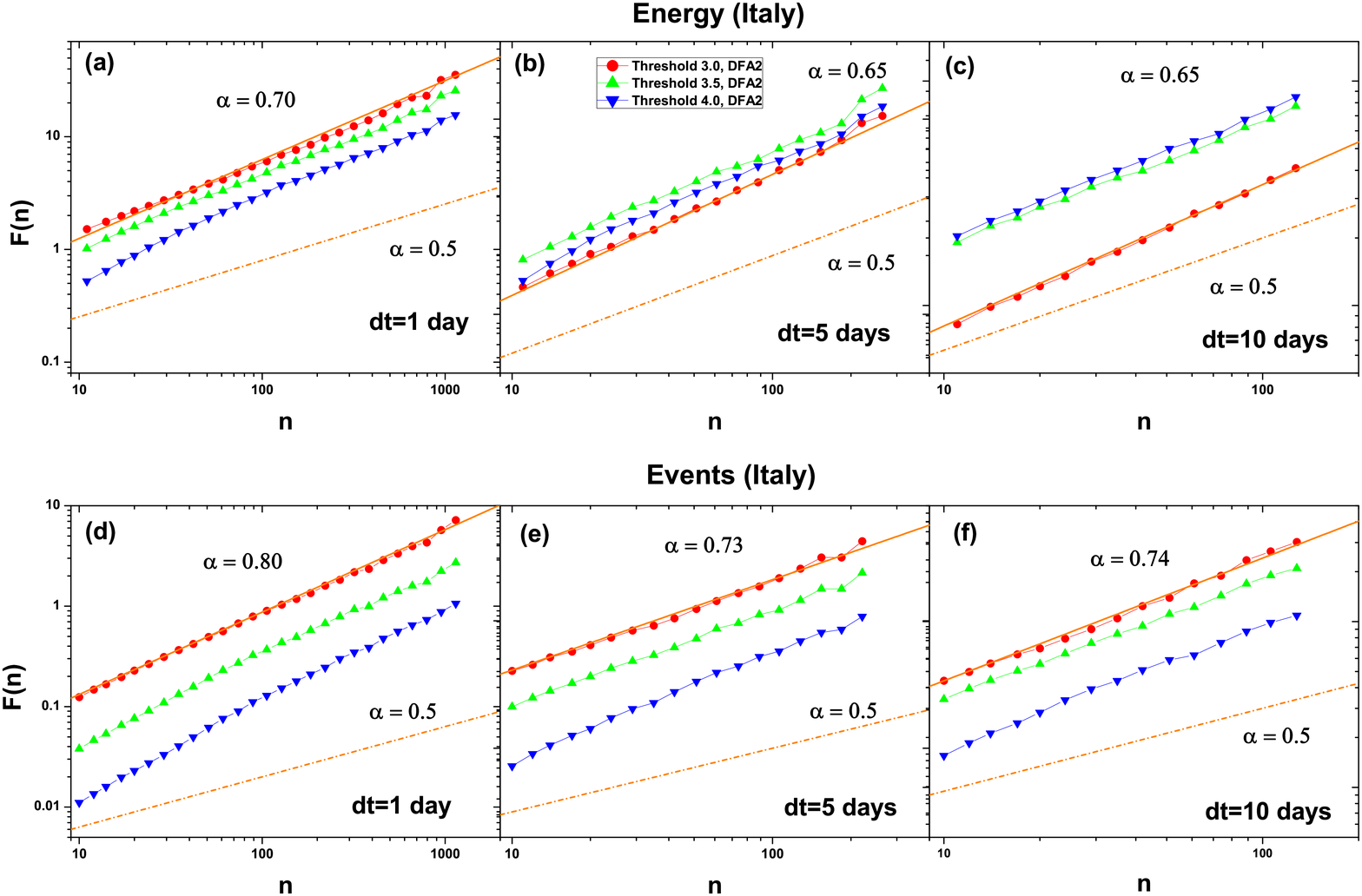}
\caption{\label{Fig:S3} DFA of (a, b and c) the released energy time series, (d, e and f) the number of events time series, defined in Eq.~(2), within a $dt$ time period for  Italian catalog. The solid lines are the best fitting lines with slope $\alpha$, R-square $>0.99$. For comparison, the dashed line with slope $\alpha=0.5$, indicating no memory, are presented.}
\end{centering}
\end{figure}

\begin{figure}
\begin{centering}
\includegraphics[width=1.0\linewidth]{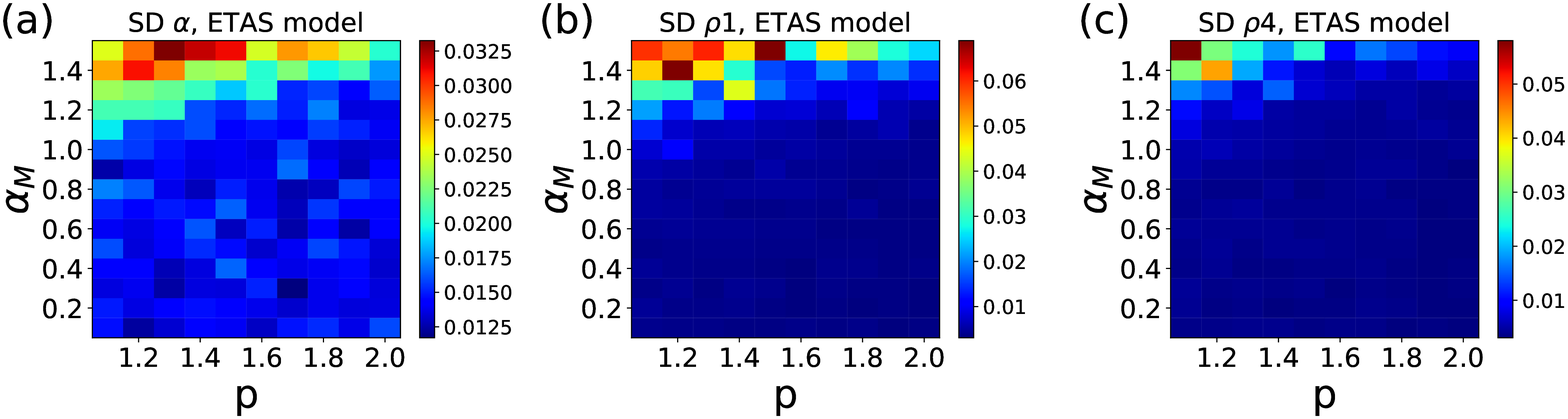}
\caption{\label{Fig:S7} The standard deviation (SD) of (a) DFA scaling exponent $\alpha$, (b) CP coefficient $\rho1$ for short recurrence times $Q1$, (c) CP coefficient $\rho4$ for long recurrence times $Q4$. The other ETAS parameters used were $A = 6.26$, $\mu = 0.2$ and $c = 0.007$.}
\end{centering}
\end{figure}

\begin{figure}
\begin{centering}
\includegraphics[width=1.0\linewidth]{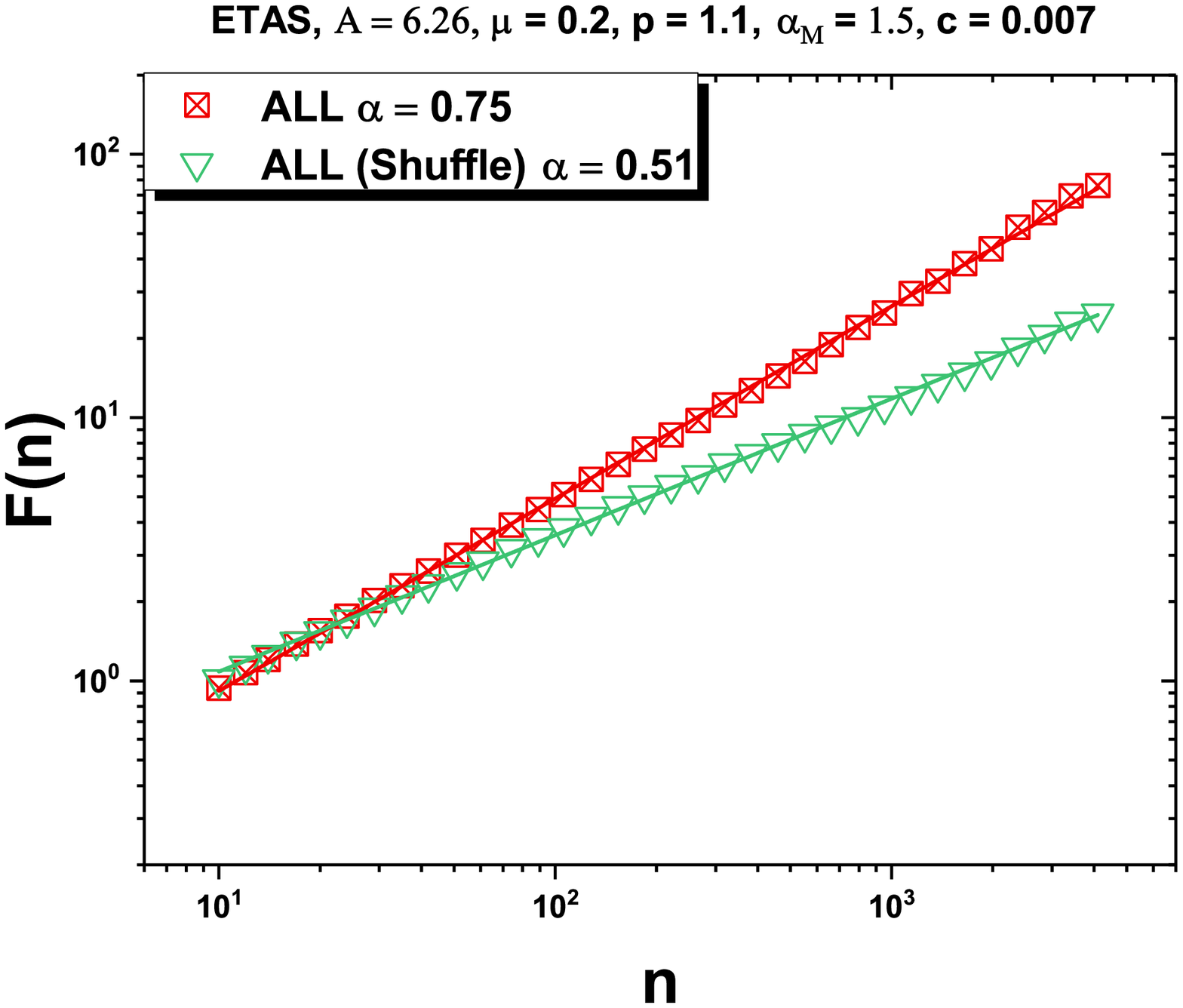}
\caption{\label{Fig:S4} Detrended fluctuation analysis of the interoccurrence times from the ETAS model, with parameters $A = 6.26$, $\mu = 0.2$, $p=1.1$, $\alpha_M = 1.5$ and $c = 0.007$. The solid line is the best fitting line with slope $\alpha = 0.75$, R-square $>0.99$. For comparison, the shuffled data with with slope $\alpha=0.5$, indicating no memory, are presented.}
\end{centering}
\end{figure}

%\begin{figure}
%\begin{centering}
%\includegraphics[width=0.75\linewidth]{Fig_S_ETAS_MC}
%\caption{\label{Fig:S5} .}
%\end{centering}
%\end{figure}

\begin{figure}
\begin{centering}
\includegraphics[width=1.0\linewidth]{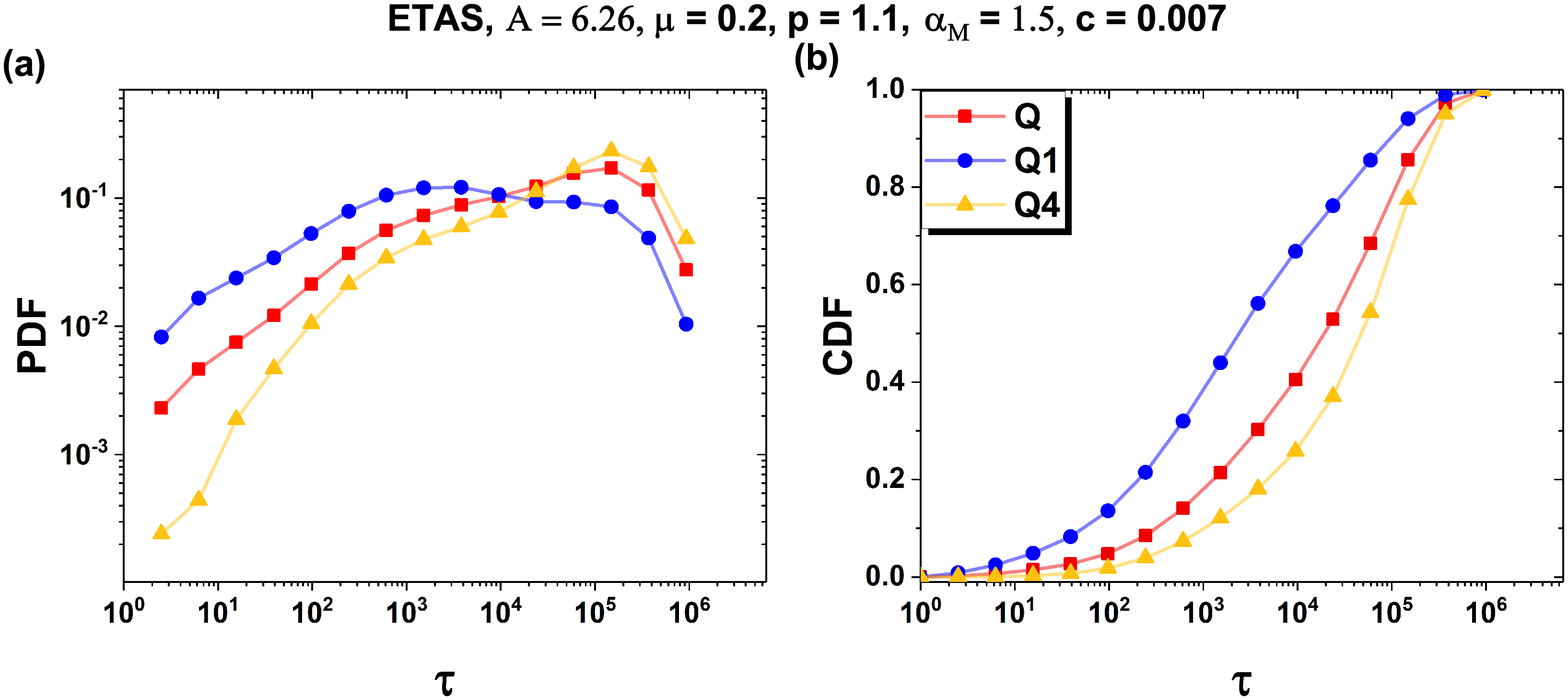}
\caption{\label{Fig:S5} Conditional (a) PDF  and (b) CDF of the recurrence times $\tau$ for the ETAS model, with parameters $A = 6.26$, $\mu = 0.2$, $p=1.1$, $\alpha_M = 1.5$ and $c = 0.007$.}
\end{centering}
\end{figure}

\begin{figure}
\begin{centering}
\includegraphics[width=1.0\linewidth]{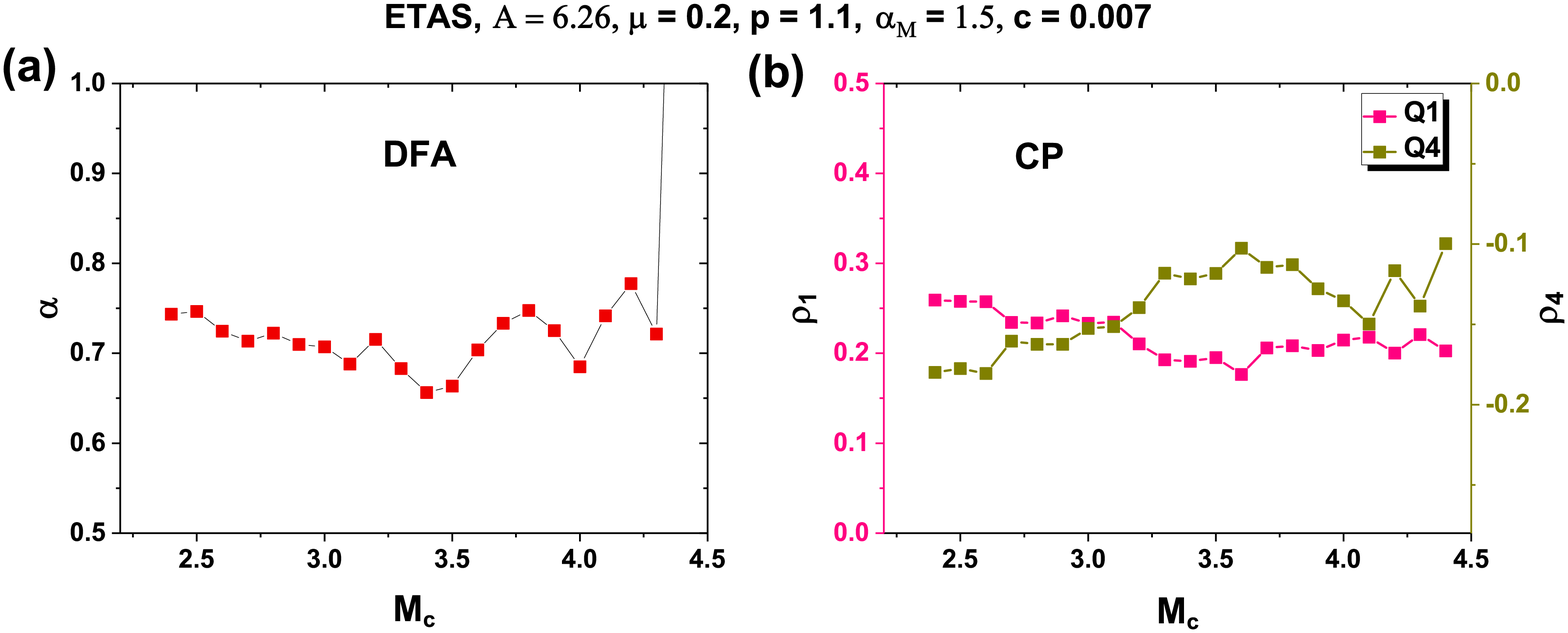}
\caption{\label{Fig:S6} (a) DFA memory scaling exponent $\alpha$ as a function of magnitude threshold $M_c$; (b) CP memory coefficient $\rho1$ and $\rho4$ as a function of magnitude threshold $M_c$ for ETAS model, with parameters $A = 6.26$, $\mu = 0.2$, $p=1.1$, $\alpha_M = 1.5$ and $c = 0.007$.}
\end{centering}
\end{figure}

%\begin{figure}
%\begin{centering}
%\includegraphics[width=1.0\linewidth]{new1}
%\caption{\label{Fig:S7} The standard deviation (SD) of (a) DFA scaling exponent $\alpha$, (b) CP coefficient $\rho$ for short recurrence times $Q1$, (c) CP coefficient $\rho$ for long recurrence times $Q4$. The other ETAS parameters used were $A = 6.26$, $\mu = 0.2$ and $c = 0.007$.}
%\end{centering}
%\end{figure}
%
%\begin{figure}
%\begin{centering}
%\includegraphics[width=1.0\linewidth]{Fig3B}
%\caption{\label{Fig:S7} The standard deviation (SD) of (a) DFA scaling exponent $\alpha$, (b) CP coefficient $\rho$ for short recurrence times $Q1$, (c) CP coefficient $\rho$ for long recurrence times $Q4$. The other ETAS parameters used were $A = 6.26$, $\mu = 0.2$ and $c = 0.007$.}
%\end{centering}
%\end{figure}

\end{document}